\documentclass[final,5p,times,twocolumn]{elsarticle}

\usepackage{amssymb,amsmath}
\usepackage{natbib}
\usepackage{color}

\journal{Journal of Theoretical Biology}

\begin{document}

\begin{frontmatter}



\title{aa-tRNA competition is crucial for the effective translation efficiency}

\author[1]{Wenjun Xia} 
\author[2]{Jinzhi Lei}

\address[1]{Zhou Pei-Yuan Center for Applied Mathematics, Tsinghua University, Beijing 100084, China}

\address[2]{MOE Key Laboratory of Bioinformatics, Zhou Pei-Yuan Center for Applied Mathematics, Tsinghua University, Beijing 100084, China}

\begin{abstract}
Translation is a central biological process by which proteins are synthesized from genetic information contained within mRNAs. 
Here we study the kinetics of translation at molecular level through a stochastic simulation model.  The model explicitly include 
RNA sequences, ribosome dynamics, tRNA pool and biochemical reactions in the translation elongation. The results show that the 
translation efficiency is mainly limited by the available ribosome number, translation initiation and the translation elongation 
time. The elongation time is log-normal distribution with mean and variance determined by both the codon saturation and the process of aa-tRNA selection at each codon binding. Moreover, our simulations show that the translation accuracy exponentially decreases with the sequence 
length. These results suggest that aa-tRNA competition is crucial for both translation elongation, translation efficiency and the 
accuracy, which in turn determined the effective protein production rate of correct proteins. Our results improve the dynamical
equation of protein production with a delay differential equation which is  dependent on sequence informations through both the effective production rate and the distribution of elongation time.

\end{abstract}


\end{frontmatter}


\section{Introduction}
\label{sec:1}
Translation is a central biological process by which genetic information contained within mRNAs is interpreted to generate  
proteins. Ribosomes provide the environment for all activities of the translation process, such as the formation of the initiation 
complex, the elongation of the translation process by which the ribosome moves along the mRNA sequence, and the dissociation of the 
ribosome from the mRNA. Protein synthesis is principally regulated at the initiation stage and hence the protein production 
rate is mainly limited by the availability of free ribosomes \cite{richard,premal}. During translations, the ribosome selects matching
aminoacylated tRNA (aa-tRNA) to the mRNA codon from a bulk of non-matching tRNAs, and the reaction rate constants can show 350-fold 
difference in the stability of cognate and near-cognate codon-anticodon complexes \cite{kirill}. Hence, in addition to the 
initiation stage, the translation efficiency is also affected by the mRNA sequence and the competition between cognate 
and near-cognate tRNAs \citep{aaron,Plotkin:2010,tamir1}. Global understanding of how ribosome number, mRNA sequence, and 
tRNA pool combine to control the translation kinetics has been an interesting topic in recent years for its potential impacts 
on the biogenesis and synthesis biology \cite{Guttman:2013,yuanhui,Ninio:2012,premal}.         

Computational models have been developed to investigate details of the translation kinetics and to explore the main factors that 
affect the translation efficiency, such as codon bias, tRNA and ribosome competition, ribosome queuing, codon 
order \cite{gina,dominique,aaron,mitarai,premal1,premal,marlena}. In these models, status of all ribosomes and tRNAs along 
a mRNA are tracked in continuous time. Translation initiation and the availability of free ribosomes were highlighted in previous 
studies \cite{dominique,premal,marlena}. In \cite{marlena}, it was found that the varying of translation efficiency were caused by very short 
times of translation initiation. Through a model that tracks all ribosomes, tRNAs and mRNAs in a cell, it was concluded 
that the protein production in healthy yeast cells was typically limited by the availability of free ribosomes, however the protein 
production under stress was rescued by reducing the initiation or elongation rates \cite{premal} . Codon bias of a mRNA sequence is an important 
factor that may affect the the translation efficiency due to competitions of tRNAs \cite{gina,dominique,aaron}.  A study 
of \textit{S. cerevisiae} genome suggest that tRNA diffusion away from the ribosome is slower than translation, and hence codon 
correlation in a sequence can accelerate translation because the same tRNA can be used by nearby codons \cite{gina}. In the 
elongation process, a cognate, near-cognate, or non-cognate tRNA may attempt to bind to the A site of a ribosome. A 
study based on a computation model with detailed tRNA pool composition shows that the competition between near-cognate and cognate 
tRNAs is a key factor that determines the translation rate \cite{aaron}.  Another study by a mean-field model of translation 
in \textit{S. cerevisiae} shows that the competition for ribosomes, rather than tRNAs, limits global translation \cite{dominique}.  
Ribosome collisions can also reduce the translation efficiency according to a model of stochastic translation process 
for \textit{E. coli lacZ} mRNA as a traffic problem \cite{mitarai}. From the point of view of evolution, the mechanism for 
controlling the efficiency of protein translation was evolutionarily conserved according to a calculation on adaptation 
between coding sequences and the tRNA pool \cite{tamir}. Moreover, using a nested model of protein translation and population 
genetics to the genomic of \textit{S. cerevisiae}, it was suggested that the codon usage bias of genes can be explained by the 
evolution through the selections for efficient ribosomal usage, genetic drift, and biased mutation, and the selection for efficient 
ribosome usage is a central force in shaping codon usage at the genomic scale \cite{premal1}.   

Despite extensive studies, much details of how translation is controlled by mRNA sequences and cellular environment remain 
exclusive.  Both the number of available free ribosomes and the codon orders are important for translation efficiency, 
however how various factors combine to determine the translation efficiency is not clearly formulated.   Since a codon is bound 
by a near-cognate tRNA, proteins with mismatched amino acids can be produced in translations.  Hence, the translation accuracy may 
be dependent on the codon usage of a sequence and the composition of tRNAs, but little result about the dependence is known to 
the best of our knowledge. The relation of how the timing of ribosome elongation stage depends on a sequence and the tRNA pool 
is closely related to the modeling of genetic network dynamics in which the elongation time associates with the time-delay in 
dynamical equations \cite{Tian:2007,Mier:2010,Zavala:2014}, but how the elongation time is formulated remains mystery. 

In this paper the translation kinetics is considered through a stochastic computation model with detailed reactions of the ribosome 
dynamics. In our study, several factors including the coding sequence, ribosomes, and the composition of tRNA tool were modeled to 
investigate how the translation efficiency, accuracy, and elongation time are determined.  Moreover, translation dynamics of 
various mRNA sequences (yeast and human, coding and non-coding mRNAs) were studied to try to clarify whether or not the 
sequence is important for the translation efficiency and the translation accuracy. Our results show that the translation 
efficiency is mainly limited  by the number of the available ribosomes, translation initiation and the elongation time of translation, 
and the elongation time is log-normal distribution with mean and variance of the logarithm of the elongation time dependent on 
the sequence through aa-tRNA usages. Moreover, the translation accuracy exponentially decrease with the sequence length. These 
results provide more detailed understanding of the translation processes, and can improve the mathematical modeling of protein 
production in gene regulation network dynamics.

\section{Model and methods}
\label{sec:2}

\subsection{Model description}
\label{sec:2.1}
\begin{figure}[htbp]
\centering
\includegraphics[scale=0.3]{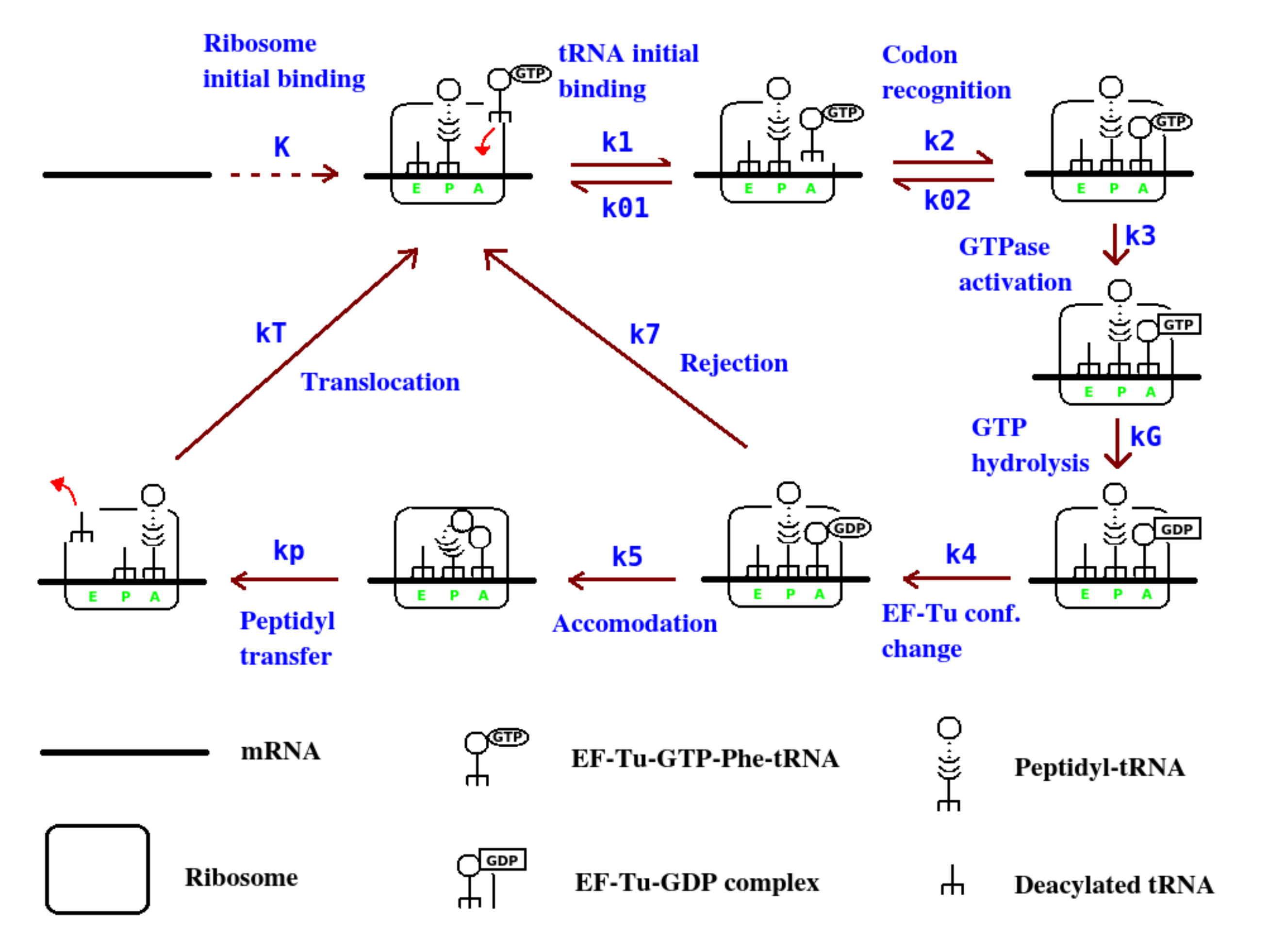}
\caption{Kinetic scheme of RNA translation. Re-drew from \cite{aaron}.}
\label{f:kinetic}
\end{figure}

Fig. \ref{f:kinetic} illustrates our model of ribosome kinetics in translation presented 
in \cite{aaron}\footnote{See \texttt{http://v.youku.com/v\_show/id\_XNzMxNzEwNjg0.html} for an animation of translation. 
Kindly provided by Prof. Ada Yonath.}.  We summarize the model description below and refer \cite{aaron} for details.

Translation of a protein begins from the initiation stage by which the start codon (AUG site) of the mRNA sequence is occupied 
by a ribosome, and the peptide between the first two amino acids are formed, with corresponding aa-tRNAs 
binding to the E and P sites of the ribosome, respectively.  During the elongation, each move of the ribosome includes 9 steps 
as shown by Fig. \ref{f:kinetic}: initial binding of aa-tRNA, codon recognition, GTPase activation, GTP hydrolysis, EF-Tu 
conformation change, rejection, accommodation, peptidyl transfer, and translocation. For each codon on the mRNA sequence, 
tRNAs in the tRNA pool are divided into three types: cognate, near-cognate, and non-cognate, as listed in \cite{aaron}. All 
aa-tRNAs can attempt to bind to the A site of the ribosome according to the match between codon and anticodon \cite{kirill}, 
however only cognate and near-cognate aa-tRNAs can go through the step of peptide formation, non-cognate aa-tRNAs are rejected 
by codon recognition. Cognate aa-tRNAs give the correct amino acid following the genetic code, while near-cognate tRNAs often 
bring incorrect amino acids and yield a defect protein. Reactions rates are different for cognate and near-cognate tRNAs, which 
have been reported at \cite{kirill,savel} and are given by Table \ref{t:k} in our simulations. We note that near-cognate 
aa-tRNAs are more likely to be rejected at both steps of codon recognition and rejection. Therewith, the competition between 
cognate and near-cognate tRNAs may be crucial for both the fidelity of peptide synthesis and translation 
efficiency \cite{aaron,kirill}.  After peptidyl transfer, the E  site aa-tRNA is released and the ribosome move forward a 
codon with the A site free waiting for the next move. Translation of a polypeptide stops when the ribosome reaches a stop 
codon (UAG/UAA/UGA), where the polypeptide is released and the ribosome drops off from the mRNA. One ribosome can synthesize 
only one polypeptide at a time, and each mRNA can be translated simultaneously by multiple ribosomes. The multiple ribosomes 
forms a queue along the mRNA, with a safe distance of at least 10 codons between two ribosomes  \cite{mitarai,premal}. 

\begin{table}[t]
\centering
\caption{Values of kinetic rate constants ($s^{-1}$) (refer to \cite{aaron})}
\resizebox{8cm}{2cm}{
\begin{tabular} {c c c c c}
\hline\hline
Parameters & Values & Cognate & Near-cognate & Non-cognate\\
[0.5ex]
\hline
K          &  0.03  &  -          &      -        &   -          \\
k1         &   -       &  140     &  140       &  2000       \\
k01       &  -        &  85       &  85         &  -          \\
k2         &   -       &  190     &  190       &  -           \\
k02       &  -        &  0.23    &  80         &  -            \\
k3         &   -       &  260     &  0.4        &  -            \\
kG        &  -        &  1000   &  1000     &  -           \\
k4         &   -       &  1000   &  1000     &  -            \\
k5         &   -       &  1000   &  60         &  -          \\
k7         &   -       &  60       &  1000     &  -          \\
kp         &   -       &  200     &  200       &  -          \\
kT         &   -       &  20       &  20         &  -          \\
\hline
\end{tabular}
}
\label{t:k}
\end{table}

\begin{table}[t]
\centering
\caption{tRNA pool composition (refer to \cite{dong,aaron}). Also refer \cite{aaron} for the anti-codons for the tRNAs.}
\resizebox{8cm}{2.5cm}{
\begin{tabular} {c c c c c c}
\hline\hline
tRNA & Molecules/cell & tRNA   & Molecules/cell & tRNA & Molecules/cell\\
[0.5ex]
\hline
Ala1 &  3250          & His    &  639           & Pro3 &  581          \\
Ala2 &  617           & Ile1   &  1737          & Sec  &  219          \\
Arg2 &  4752          & Ile2   &  1737          & Ser1 &  1296       \\
Arg3 &  639           & Leu1   &  4470          & Ser2 &  344      \\
Arg4 &  867           & Leu2   &  943           & Ser3 &  1408       \\
Arg5 &  420           & Leu3   &  666           & Ser5 &  764       \\
Asn  &  1193          & Leu4   &  1913          & Thr1 &  104       \\
Asp1 &  2396          & Leu5   &  1031          & Thr2 &  541        \\
Cys  &  1587          & Lys    &  1924          & Thr3 &  1095     \\
Gln1 &  764           & Met f1 &  1211          & Thr4 &  916     \\
Gln2 &  881           & Met f2 &  715           & Trp  &  943     \\
Glu2 &  4717          & Met m  &  706           & Tyr1 &  769      \\
Gly1 &  1068          & Phe    &  1037          & Tyr2 &  1261     \\
Gly2 &  1068          & Pro1   &  900           & Val1 &  3840    \\
Gly3 &  4359          & Pro2   &  720           & Val2A&  630  \\
     &                &        &                & Val2B&  635  \\
\hline
\end{tabular}
}
\label{t:t}
\end{table}

\subsection{Numerical scheme}
The translation process with multiple ribosomes was modeled with the stochastic simulation algorithm (SSA) \cite{gillespie}, which 
includes the following reactions:
\begin{enumerate}
 \item binding of a ribosome to the start codon if the first 10 codons are not occupied by ribosomes;
 \item binding of an aa-tRNA from the tRNA pool to the A site of an unoccupied ribosome;
 \item reactions of codon recognition, energy transformation, and peptide formation;
 \item releasing of the tRNA from the E site of a ribosome;
 \item translocation of the ribosome to the next codons if the safety condition is satisfied;
 \item dropping off of the ribosome once the stop codon is reached. 
\end{enumerate}

Kinetics parameters are given by Table \ref{t:k}, which refer to \cite{aaron}. The tRNA pool compositions referred the total number 
of each tRNA in a yeast cell from \cite{dong, aaron} and are given by Table \ref{t:t}.  In simulations, to mimic the effects of 
available tRNAs for each single mRNA translation, we used a factor $F$ ($0<F\leq 1$) to all tRNA numbers to adjust the changes in the numbers of 
available tRNAs. For the anti-codon of each tRNA and the cognate, near-cognate, and non-cognate for each codon, refer \cite{aaron} 
for details. 

It has been shown that the availability of free ribosomes is an important limitation for the translation efficiency \cite{premal}.  
Here we introduced a parameter $R$ for the maximum number of available ribosomes can be used for a single sequence translation.  
We note that a ribosome can be re-used after it was released from the stop codon. 

\begin{figure}[htbp]
\centering
\includegraphics[width=8cm]{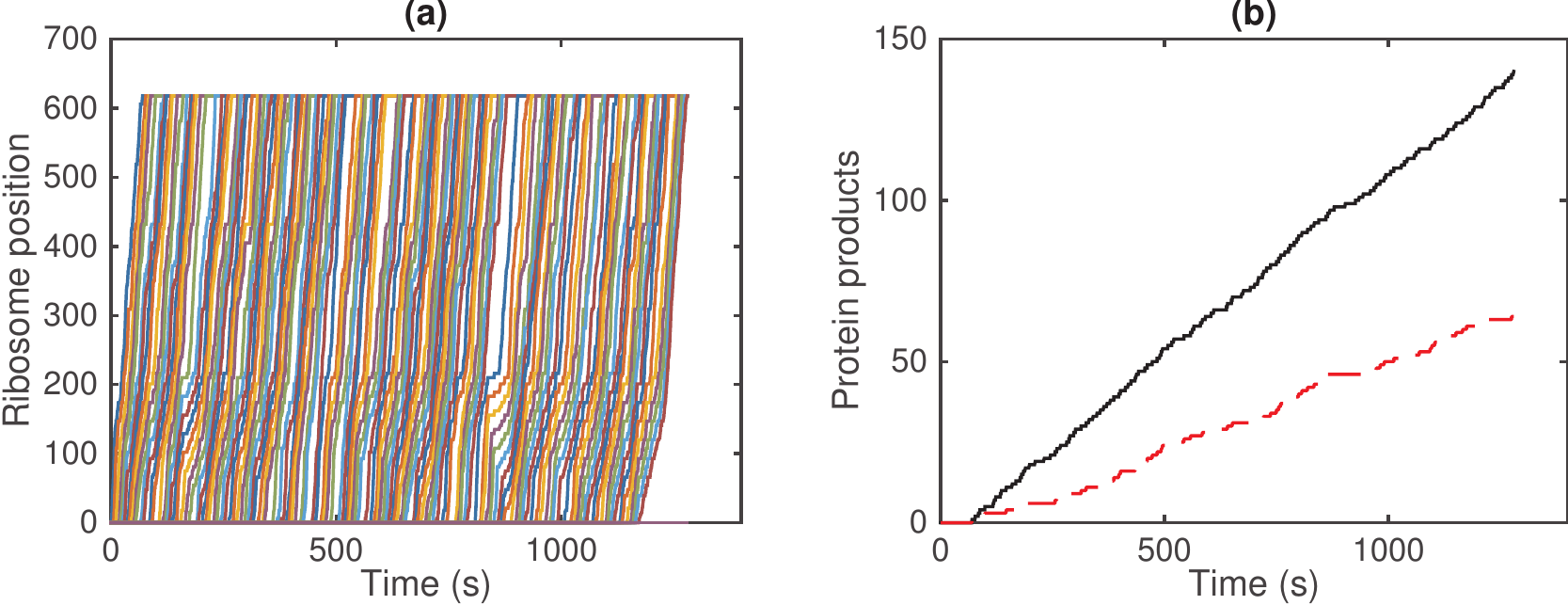}

\caption{Translation kinetics of a single mRNA sequence. (a) Positions of each ribosome on the sequence.   (b) Numbers of protein 
products. Black solid line for all protein products, red dashed line to correctly translated proteins (no incorrect amino acid
added by near-cognate aa-tRNAs). Here the sample sequence is the gene YAL003W from the SGD yeast coding sequences, with sequence 
length $L = 621 nt$. Parameters are $R=20$, $F=0.03$ and other parameters refer Table \ref{t:k}. }
\label{f:ribtraj}
\end{figure}

An example of translation kinetics obtained from our simulation is given at Fig. \ref{f:ribtraj}, which shows that the ribosomes 
sequencing along the mRNA, and the number of protein production increases linearly with the translation time.  The average 
translation rate (amino acids per second) in our simulation is of a order of $10$, in good agreement with the experimental 
observations \cite{proshkin}. These suggest well defined translation efficiency, elongation time, and accuracy of a translation 
as given below. 

\subsection{Translation efficiency, elongation time, and accuracy}

To quantify the translation process, we consider the translation efficiency for the protein production rate, the elongation 
time for move kinetics of each individual ribosome, and the translation accuracy for the fidelity of translation.  
The \textit{translation efficiency} ($TE$) is defined as the average slope of the increasing of protein production number 
with the translation time. The \textit{elongation time} of each ribosome is given by the time period from the binding of a 
ribosome to the start codon to its dropping off from a stop codon. The \textit{elongation time per codon} ($ETC$), the average 
time for a ribosome to move one codon, is often used to describe the translation kinetics. The elongation time is given 
by $ETC\times L/3$, where $L$ is the length (in \textit{nt}) of a mRNA sequence. Since a protein product may contain 
mismatched amino acids due to the binding of near-cognate aa-tRNAs with the mRNA,  it was possible to have incorrect protein products 
in the translation. Hence, the ratio of correct proteins in all protein produces gives the \textit{translation accuracy}.     

\section{Results}
\label{sec:3}

\subsection{Translation elongation time is log-normal distribution and sequence dependent}

The elongation time measures how long it takes a ribosome to finish the translation of a protein, which corresponds to the 
delay of translation in modeling the dynamics of gene regulation networks through delay differential 
equations \cite{Mier:2010,Zavala:2014}. The production of proteins can be described by translation efficiency $\alpha$ and 
mRNA number $M(t)$ through a delay differential equation of form
\begin{equation}
\label{eq:P}
\dfrac{d P}{d t} = \alpha \int_0^{+\infty} M(t-\tau) \rho(\tau) d \tau,
\end{equation}
where $\tau$ represents the elongation time, with distribution density $\rho(\tau)$. 

\begin{figure}[htbp]
\centering
\includegraphics[width=6cm]{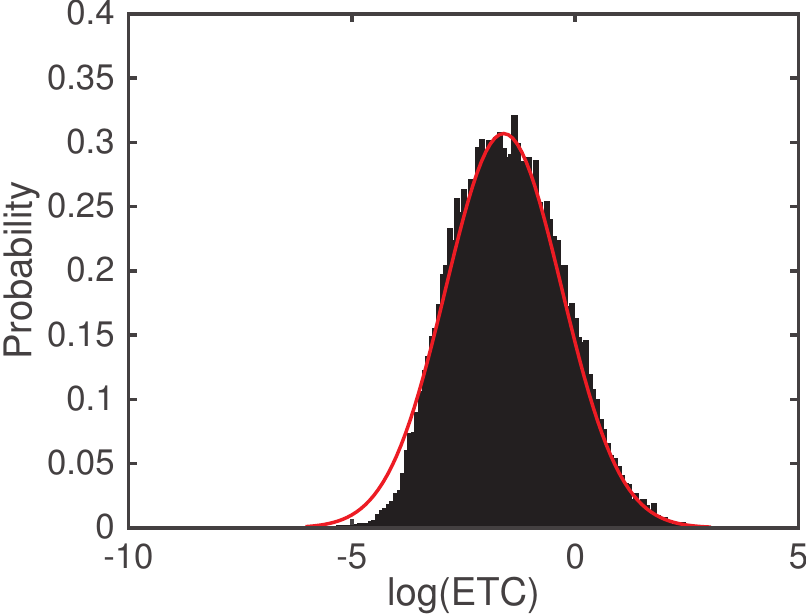}
\caption{Distribution of the elongation time per codon in the translation of YAL003W. All parameters are the same at 
Fig. \ref{f:ribtraj}. Red curve is the fit with normal distribution $\ln \mathcal{N}(-1.6, 1.69)$.}
\label{f:elong1}
\end{figure}

To obtain the formulation of the distribution density $\rho(\tau)$, we calculated the elongation time per codon ($ETC)$ in 
the translation of YAL003W (here we note $\tau = ETC \times L/3$). The distribution density is showed at Fig \ref{f:elong1}.  
The density function was well fitted by log-normal distribution
\begin{equation}
\label{eq:ga}
\ln \mathcal{N}(\mu, \sigma^2) = \dfrac{1}{x\sigma\sqrt{2\pi}}e^{-\frac{(\ln x - \mu)^2}{2\sigma^2}},\quad x>0.
\end{equation}
Here the shape parameters $\mu$ and $\sigma$ are taken so that the logarithm of \textit{ETC} has mean $\mu$ and 
variance $\sigma^2$. Let $n = L/3$ be the number of amino acids in a protein product, the density function $\rho(\tau)$ of 
elongation time is given from the log-normal distribution Eq. \ref{eq:ga} as
\begin{equation}
\label{eq:rho}
\rho(\tau) = \dfrac{1}{\tau \sigma \sqrt{2\pi}} e^{-\frac{(\ln \tau - \ln n - \mu)^2}{2\sigma^2}},
\end{equation}
and the average elongation time is 
\begin{equation}
\label{eq:bt}
\bar{\tau} = \int_0^{+\infty} \tau \rho(\tau) d \tau  = n e^{\mu + \sigma^2/2}.
\end{equation}
In the next section we show that the translation efficiency is dependent on the average elongation time, which refines our 
dynamical equation for protein production.

\vspace{0.5cm}

Each move of a ribosome consists of several chemical reactions shown by Fig. \ref{f:kinetic}, including selections of cognate 
or near-cognate aa-tRNA from the tRNA pool, and a step forward if the safety condition is satisfied. To investigate how 
the $ETC$ depends on the mRNA sequence and translation kinetics, we examined translations of a set of 1000 sequences from 
yeast coding genes with length $L$ varies from $51$ to $1995$ \textit{nt} ($17$ to $665$ codons). To measure the tRNA usage 
of each sequence, we calculated the average fraction of cognate, near-cognate, and non-cognate tRNA along the sequence, which 
are defined as
\begin{equation}
F_\nu= \dfrac{1}{L/3}\sum_{i=1}^{L/3} \frac{n_{i,\nu}}{\mathrm{Total\ tRNA\ number}},\ \nu = \mathrm{cog,\ near,\ non},
\end{equation}
here $F_{\nu}$ ($\nu = \mathrm{cog,\ near,\ non}$) measures the average tRNA usage of cognate, near-cognate, and non-cognate 
tRNAs, respectively. The summation is taken over all codons, and $n_{i,\nu}$ is the number of tRNAs of type $\nu$ for 
codon $i$ along the mRNA sequence.  

\begin{figure}[htbp]
\centering
\includegraphics[width=8cm]{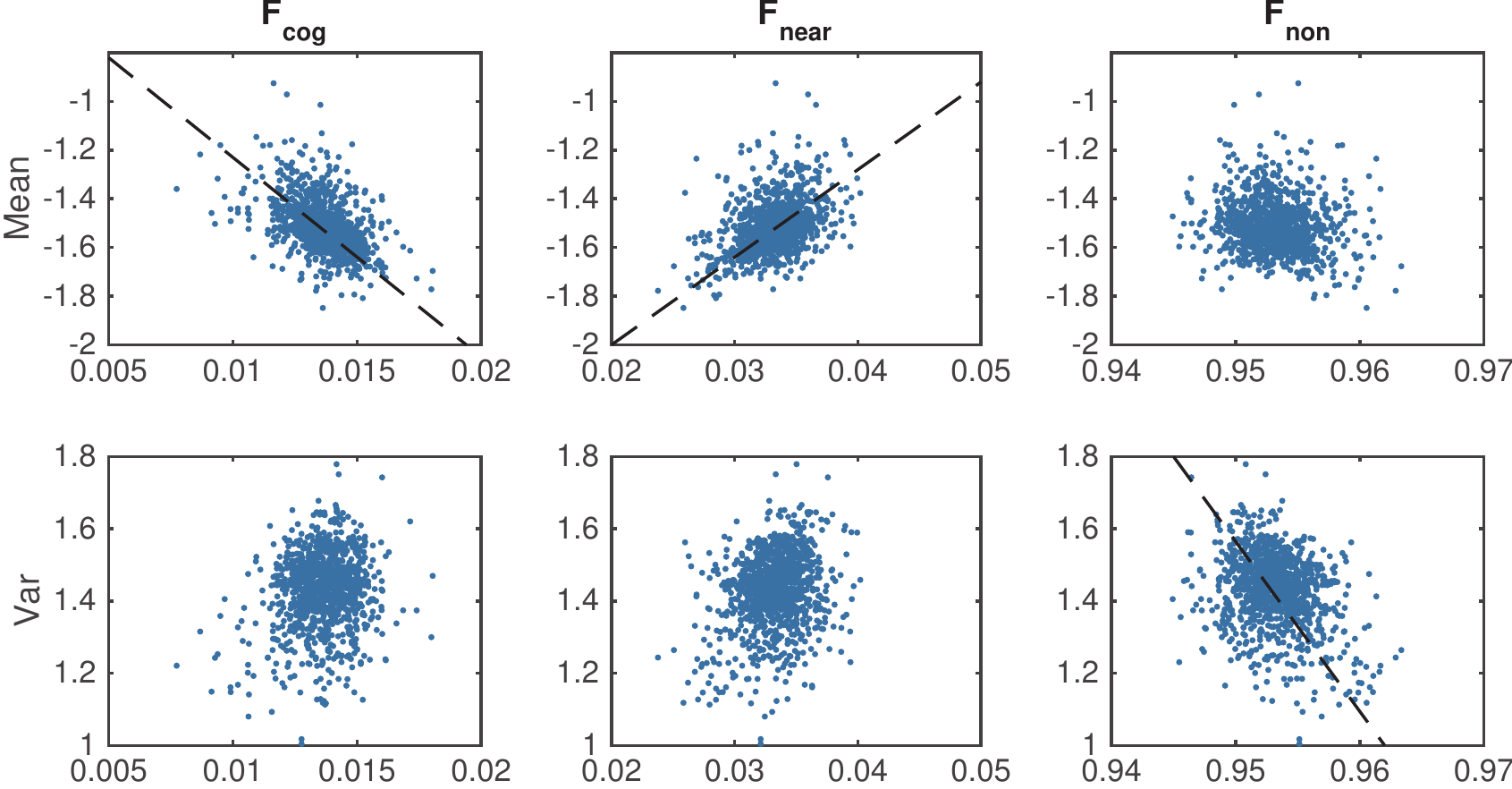}
\caption{Dependence of the $ETC$ of yeast coding sequences on tRNA usages. Dots show the mean (upper panel) and variance (bottom panel) of the logarithm of $ETC$ with cognate tRNA usage $F_{\mathrm{cog}}$, near-cognate tRNA usage $F_{\mathrm{near}}$ and non-cognate tRNA usage $F_{\mathrm{non}}$, respectively. Dashed lines show the linear fitting. Simulations of 1000 yeast coding sequences are shown, each dot corresponds to one sequence. All parameters are the same as Fig. \ref{f:ribtraj}.}
\label{f:elong2}
\end{figure}

Fig. \ref{f:elong2} shows the dependence of the mean ($\mu$) and variance ($\sigma^2$) of the logarithm of $ETC$ with tRNA 
usages. Results suggest that the mean decreases with the cognate tRNA usage $F_{\mathrm{cog}}$, increases with the 
near-cognate tRNA usage $F_{\mathrm{near}}$, and has no correlation with the non-cognate tRNA usage $F_{\mathrm{non}}$, 
while the variance is not dependent on either $F_{\mathrm{cog}}$ or $F_{\mathrm{near}}$, but weakly  decreases 
with $F_{\mathrm{non}}$. These results suggest that the competition of near-cognate tRNAs tends to increase the elongation 
time, while the competition of non-cognate tRNAs has only little affects to the elongation time. Moreover, Fig. \ref{f:elong2} 
suggests typical parameters for the distribution of the ETC of yeast coding gene translation 
are $\mu \approx -1.5$ and $\sigma^2 \approx 1.4$ (refer to Eq. \ref{eq:ga}).  Our simulations suggest no obvious dependence 
of $ETC$ with sequence length $L$ (data not shown).
 
\begin{figure}[htbp]
\centering
\includegraphics[width=8cm]{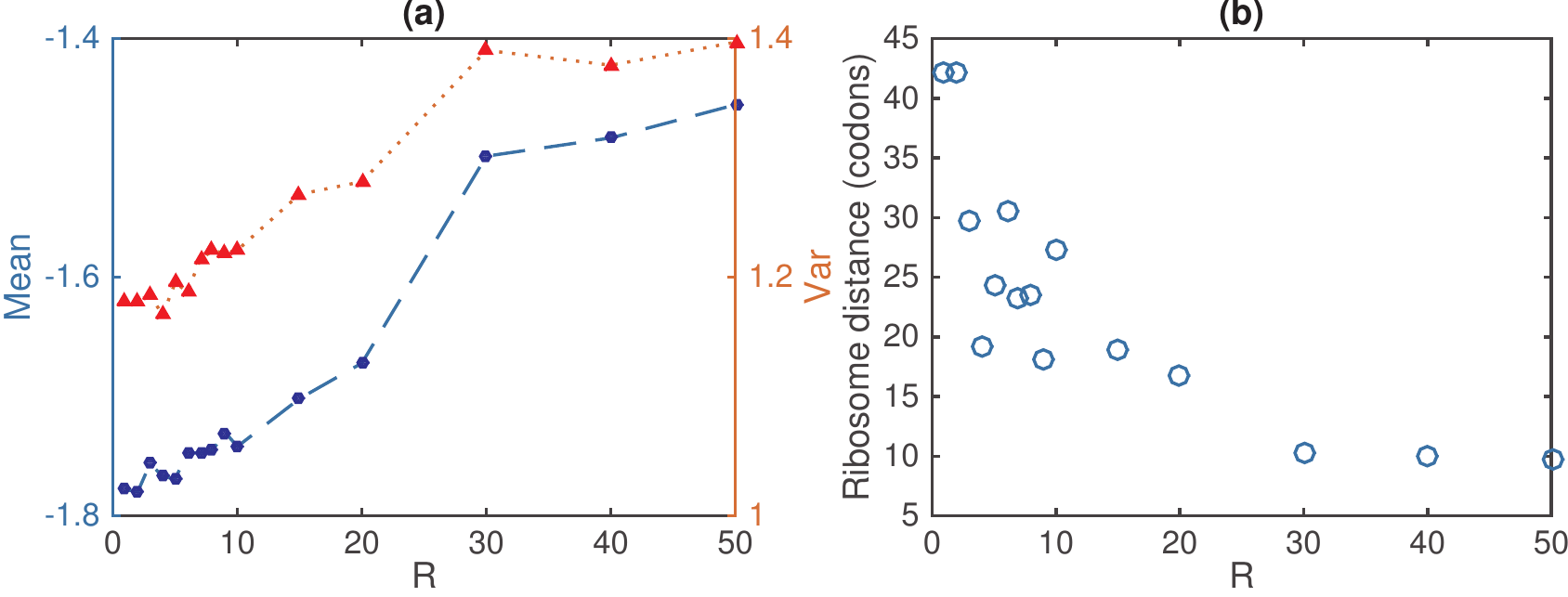}
\caption{Dependence of the elongation time with the available ribosomes number $R$. (a) Average $ETC$ versus $R$. (b) Ribosome distance (in codons) versus $R$. Sequence and parameters are the same as Fig. \ref{f:ribtraj}.}
\label{fig:elR}
\end{figure}

The investigate how the available ribosomes number $R$  affects the elongation time, we changed the value $R$ to calculate the dependence of $ETC$. Results showed that both mean and variance of the logarithm of $ETC$ are dependent on $R$ nonlinearly: mostly independent to $R$ when $R$ is either small or large, and an obvious increasing dependence when $R$ takes intermediate values (Fig. \ref{fig:elR}a). A possible reason for the increasing of $ETC$ is the traffic jam due to codon occupation. Fig. \ref{fig:elR}b shows that the average ribosome distance obviously decreases with $R$ at the intermediate region, and approaches a minimum distance (the safe distance of 10 codons) when $R$ is large. These results reveal that the increasing dependence of the elongation time with ribosome number $R$ ($10 < R < 30$) is due to the increasing of traffics jam in translation kinetics. 

\begin{figure}[htbp]
\centering
\includegraphics[width=6cm]{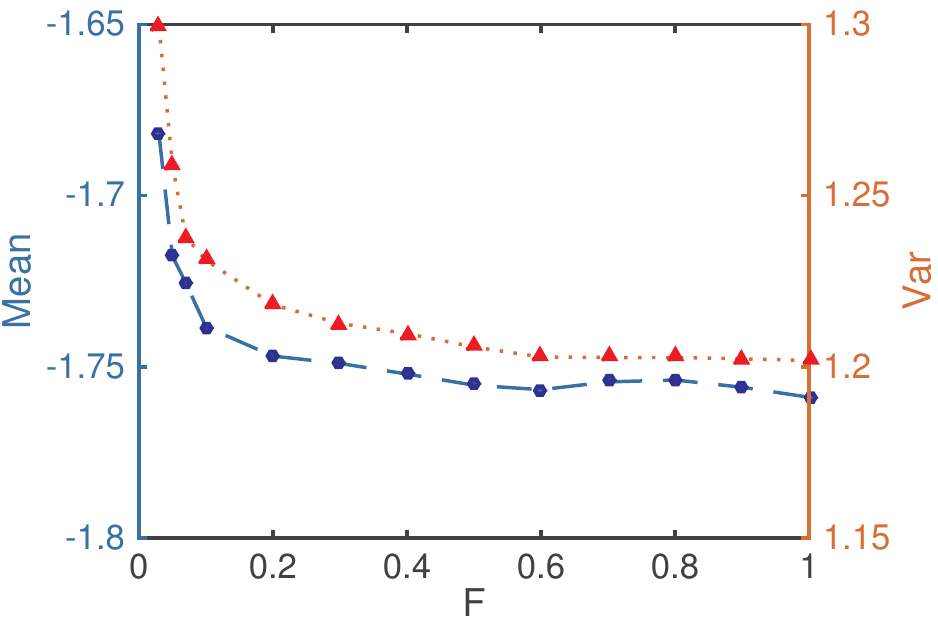}
\caption{Dependence of the $ETC$ with total tRNAs number represented by the factor $F$. The mean (left hand ordinate, blue circles 
connected with a dashed line) and variance (right hand ordinate, red triangles connected with a dotted line) of the logarithm 
of $ETC$ are shown as a function of the factor $F$.  Sequence and parameters are the same as Fig. \ref{f:ribtraj}.}
\label{fig:elf}
\end{figure}

In the above calculations, the total number of tRNAs was fixed. To further examine how the number of total tRNAs affects the 
elongation time, we varied the factor $F$ from $0.03$ to $1$ to calculate the dependence of $ETC$. Results showed that both 
mean ($\mu$) and variance ($\sigma^2$) of the logarithm of $ETC$ are decreasing with $F$ for small $F$, and nearly unchanged 
when $F > 0.5$ (Fig. \ref{fig:elf}). Biologically these dependences are obvious because it takes longer time to select a 
cognate or near-cognate when there are no enough tRNAs.

\vspace{0.5cm}

\subsection{Translation efficiency is mainly dependent on the elongation time and available ribosomes number}

\label{ss:expo}

In \cite{premal,marlena}, it has been shown that the varying of translation efficiency were caused by translation initiation, and the 
availability of free ribosomes was a typical rate limit of translation. To investigate how the translation efficiency 
depends on the translation kinetics and mRNA sequences, we constructed a model to track the dynamics of available ribosomes.  

Consider a mRNA with $n$ codons ($n = L/3$). Let $R$ be the number of available ribosomes, $x_i(t)\ (i=1,\cdots, n)$ the number 
of ribosomes at the $i^{\mathrm{th}}$ codon at time $t$, and $x_0(t)$ the number of free ribosomes.  Kinetics of a ribosome in 
translation is a combination of initiation in a rate $K$, elongation per codon in a rate $c$  and termination in a rate $K_T$. Therefore, the 
dynamics of $x_i$ can be expressed by the following differential equations model
\begin{equation}
\label{eq:rib}
\left\{
\begin{array}{rcl}
\dfrac{d x_0}{d t}&=&K_T x_n - K x_0\\
\dfrac{d x_1}{d t} &=& K x_0 - c x_1\\
\dfrac{d x_i}{d t} &=& c (x_{i-1} - x_i)\qquad i = 2,3,\cdots, n-1\\
\dfrac{d x_n}{d t}&=& c x_{n-1} - K_T x_n. 
\end{array}
\right.
\end{equation} 
The protein production rate is proportional to $x_n$. Here we note 
\begin{equation}
0\leq x_0\leq R, \quad 0\leq x_i \leq 1\quad (i=1,2,\cdots,n),
\end{equation} 
and 
\begin{equation}
\sum_{i=0}^n x_i = R.
\end{equation}

When $K_T > c$ and $R$ is small, Eq. \ref{eq:rib} has a stable equilibrium state which gives
\begin{equation}
\label{eq:xn}
x_n = \dfrac{R}{K_T \dfrac{n-1}{c} + 1 + \dfrac{K_T}{K}}.
\end{equation}
Hence, let $\bar{\tau} = \dfrac{n-1}{c}$ be approximate to the elongation time (here we note $1/c$ corresponds to the average of \textit{ETC}), the translation efficiency satisfies 
\begin{equation}
\label{eq:TE0}
TE\propto \dfrac{R K}{K \bar{\tau} +1 + K/K_T}.
\end{equation}

\begin{figure}[htbp]
\centering
\includegraphics[width=6cm]{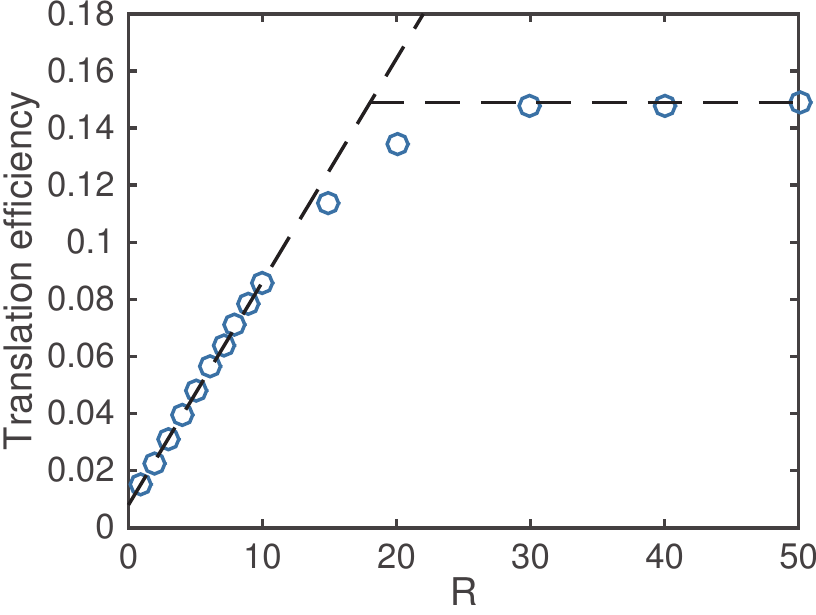}
\caption{Dependence of translation efficiency with the maximum number of available ribosomes $R$. Dashed lines show the two-phase dependence following Eq. \ref{eq:TE}. Sequence and parameters are the same as Fig. \ref{f:ribtraj}.}
\label{fig:teR}
\end{figure}

When $R$ is large so that all codons are occupied, the translation efficiency is mainly determined by the elongation time so that $TE\propto 1/\bar{\tau}$. Hence, taking account Eq. \ref{eq:TE0}, the translation efficiency can be approximated as
\begin{equation}
\label{eq:TE}
TE \propto \dfrac{K\,\min\{R,R_{\max}\}}{K \bar{\tau} +1 + K/K_T},
\end{equation}
where $R_{\max}$ is the number of available ribosomes to saturate all codons. We take $R_{\max} = n/10$ in our simulations, which is consistent with Fig. \ref{fig:elR}. We note that $\bar{\tau}$ is dependent on $R$ according to the above discussions, hence the relation Eq. \ref{eq:TE} suggests the following dependence of translation efficiency on the ribosome number $R$: linearly increases when $R$ is small, independent to $R$ when $R$ is large, and nonlinear dependence through the elongation time $\bar{\tau}$ when $R$ takes intermediate values. These results are in agree with our numerical simulations (Fig. \ref{fig:teR}).

\vspace{0.5cm}

The result Eq. \ref{eq:TE} supports the previous findings that translation initiation and ribosome number are rate limits of protein 
production. Moreover, the translation efficiency decreases with the elongation time, which shows the dependence of protein 
production with the mRNA sequence through the elongation dynamics.    
 
\begin{figure}[htbp]
\centering
\includegraphics[width=8cm]{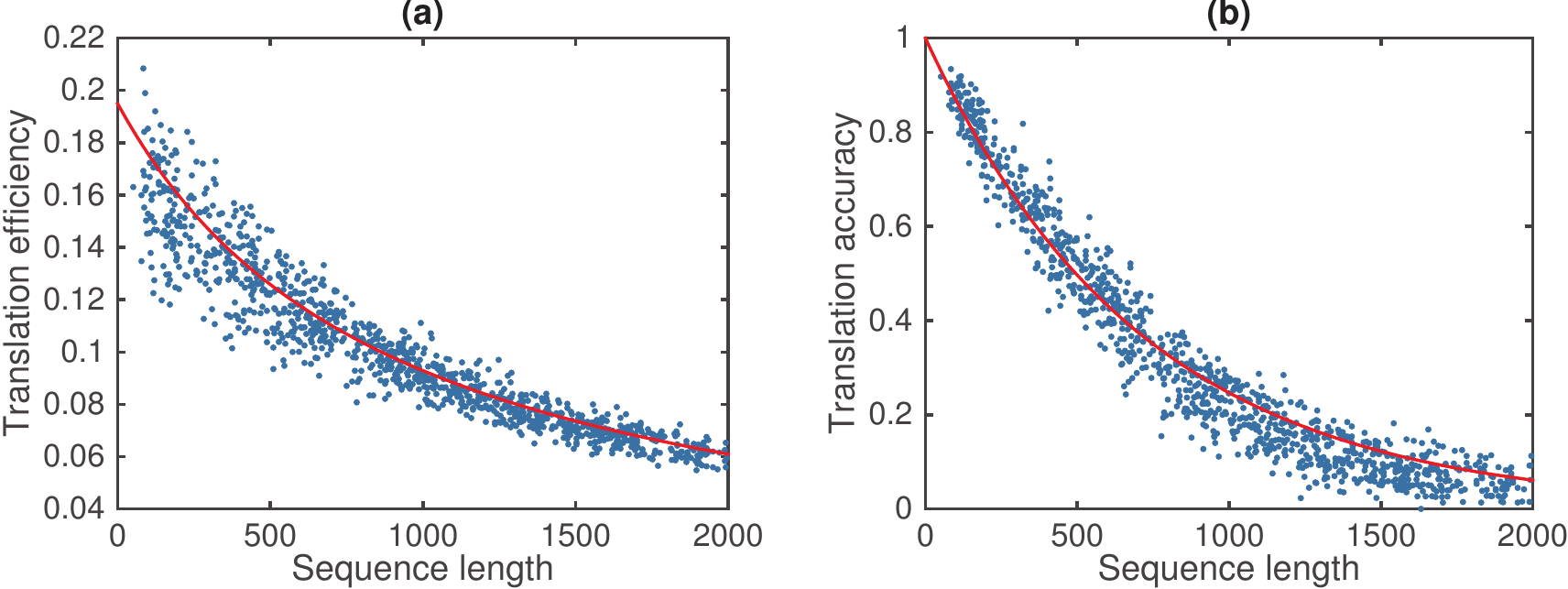}
\caption{Translation kinetics. (a) Translation efficiency versus sequence length for 1000 yeast coding sequences. Red line 
shows the fitting with $TE = \dfrac{0.195}{1+0.0033 n}$. (b) Translation accuracy versus sequence length for 1000 yeast 
coding genes. Red line shows the fitting with $e^{-0.0042 n}$. Here $n=L/3$ is the protein chain length. Data obtained 
from the simulation at Fig. \ref{f:elong2}. }
\label{f:averET}
\end{figure}

Since the average elongation time $\bar{\tau}$ is proportional to the protein length $n$, Eq. \ref{eq:TE} suggests that the translation 
efficiency depends on the protein length $n$ through a Michaelis-Menten function. Fig. \ref{f:averET}a shows translation 
efficiency versus sequence length for yeast coding sequences with different length. The translation efficiency is well fitted 
by a Michaelis-Menten function, in agreement with our theoretical conclusion Eq. \ref{eq:TE}. 

To further investigate the sensitivity of translation efficiency with the changes in parameters, we increased or decreased each 
of the parameters in Table \ref{t:k} and examined the changes in the translation efficiency. The results showed that the 
translation efficiency is sensitive with the changes in $\mathrm{k02}$ (or $\mathrm{ke02}$ for near-cognate tRNA), 
$\mathrm{k01}$, $\mathrm{k1}$, $\mathrm{k2}$, which correspond to the process of aa-tRNA selection. The translation 
initiation $K$ is also important for the translation efficiency, as we have seen from Eq. \ref{eq:TE}. Changes in other 
parameters led to minor changes in the translation efficiency. These results indicate that the steps of aa-tRNA selection 
are crucial for the translation efficiency through their effects to the elongation time, and changes in the steps of 
peptide formation have minor effects to the translation efficiency. 

\begin{figure}[htbp]
\centering
\includegraphics[width=8cm]{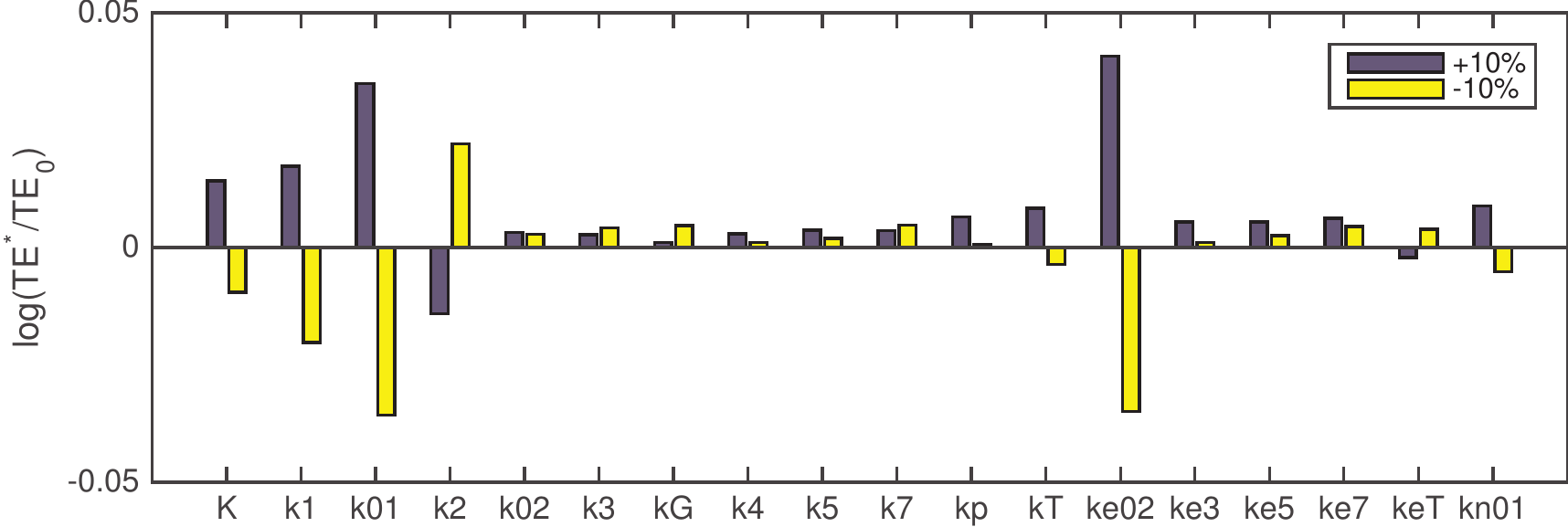}
\caption{Sensitivity analysis of the translation efficiency. Bars show changes in the logarithm of translation efficiencies induced by changes in a single parameter $\ln(TE^*/TE_0)$, where $TE^*$ and $TE_0$ represents the $TE$ for modified and default parameters, respectively. Blue bars correspond to increasing of a parameter by $10\%$, and yellow bars correspond to decreasing of a parameter by $10\%$. For parameters refer to Table \ref{t:k}, the parameters $\mathrm{ke02}$, $\mathrm{ke3}$, $\mathrm{ke5}$, $\mathrm{ke7}$, $\mathrm{keT}$ for values of $\mathrm{k02}$, $\mathrm{k3}$, $\mathrm{k5}$, $\mathrm{k7}$, $\mathrm{kT}$ of near-cognate tRNAs (second column in Table \ref{t:k}), respectively, and $\mathrm{kn01}$ for the parameter $\mathrm{k01}$ for non-cognate tRNAs (third column in Table \ref{t:k}.  Sequence and defaults parameters are same as Fig. \ref{f:ribtraj}. }
\label{fig:sen}
\end{figure}

\subsection{Translation accuracy decreases exponentially with sequence lengths}

In translations, protein products may contain mismatched amino acids  when a near-cognate aa-tRNA is selected and 
successfully form a peptide. Hence the translation accuracy, fraction of correct protein products, should be exponentially 
decay with the chain length, the decay rate is associated with the probability of selecting a near-cognate aa-tRNA at 
each step. Fig. \ref{f:averET}b shows the translation accuracy versus sequence length, which is well fitted with an 
exponential function.

In living cells, abnormal proteins are usually degraded quickly so that the intracellular amino acids can be recycled 
efficiently. Hence, only correctly translated proteins are relevant in modeling the dynamics of gene regulation networks. 
This yields a factor by translation accuracy in the production rate of normal proteins in the equation Eq. \ref{eq:P}. 

The above discussions suggest a more refined equation for effective protein production Eq. \ref{eq:P} with $\rho(\tau)$ 
given by log-normal distribution Eq. \ref{eq:rho}, and the effective translation efficiency $\alpha$ is given by
\begin{equation}
\label{eq:alpha}
\alpha = \dfrac{a e^{-c n}}{1 + b n e^{\mu + \sigma^2/2}},
\end{equation}
where parameters $a, b$ depend on available ribosomes number, translation initiation and termination, and $c$ 
relates with the composition of the tRNA pool. A crucial refinement of Eq. \ref{eq:alpha} is the 
dependence of protein chain length $n$, and other parameters are somehow universal for differential proteins, except 
weak dependence of $\mu$ and $\sigma^2$ with sequences as shown by Fig. \ref{f:elong2}, under certain cellular conditions.

\subsection{Translation kinetics with sequence dependences}

\begin{figure}[htbp]
\centering
\includegraphics[width=8cm]{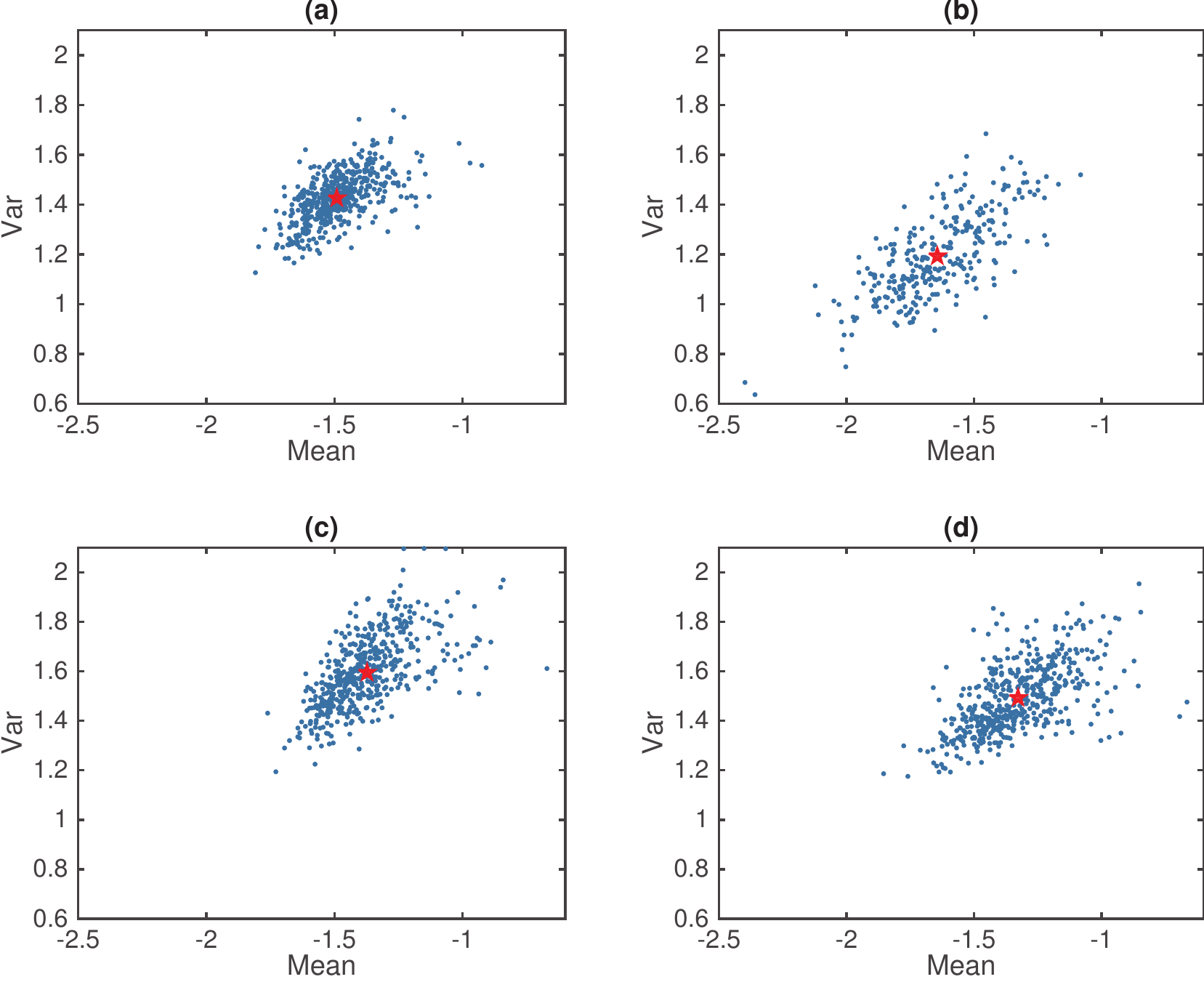}
\begin{center}
\begin{tabular}{l|cc}
\hline
& Mean ($\mu$) & Variance ($\sigma^2$) \\
\hline
Yeast coding 		& $-1.4895$ 	& $1.4256$ \\
Yeast non-coding 	& $-1.6410$	& $1.1948$ \\
Human coding		& $-1.3773$	& $1.5981$ \\
Human non-coding	& $-1.3295$	& $1.4885$ \\
\hline
\end{tabular}
\end{center}
\caption{$ETC$ of the translations for different samples. Distributions of Mean and variance of the logarithm of 
$ETC$ for yeast coding RNAs (a), yeast noncoding RNAs (b), human coding RNAs (c) and human noncoding RNAs (d). 
Here the results of 500 random sequences with length $200 nt < L< 1000 nt$ for each sample are shown. Red stars show 
the average values for each sample, values are given by the table. Parameters are $R=20, F=0.03$, and other parameters are referred to Table \ref{t:k}.}
\label{f:cnchuman}
\end{figure}

A motivation of this study was trying to examine whether there are distinct dynamics for coding and non-coding RNA 
sequences in the translation. We have shown that the translation efficiency depends on the mRNA sequence through the 
elongation time, and the mean and variance of the elongation time per codon are dependent on the sequence through 
the aa-tRNA usages. A study of ribosome occupancy showed that many large noncoding RNAs are bound by ribosomes and 
hence are possible to be translated into proteins \cite{nicholas, Ingolia:2011}. To investigate the translation 
kinetics of coding and noncoding RNAs, we applied the model simulation to yeast coding RNA, yeast noncoding RNA, human 
coding RNA and human noncoding RNA. In each sample, 500 sequences with lengths between $200 nt$ and $1000 nt$ were 
selected, however most of the noncoding RNA has reading frames with lengths less than $300 nt$, in agreement with 
the observations in \cite{naturegenetics}.  Simulations showed that the previous results are qualitatively 
held for different samples. Fig. \ref{f:cnchuman} shows the distributions of mean ($\mu$) and variance ($\sigma^2$) 
of the logarithm of the $ETC$ for each set of the simulations (The average of $\mu$ and $\sigma^2$ for each ample are given by the table), which are crucial parameters in the density function 
Eq. \ref{eq:ga}.  From Fig. \ref{f:cnchuman}, we have the following observations:  coding RNAs (both yeast and human) 
have similar distribution in $\mu$ and $\sigma^2$; noncoding RNAs have smaller variance $\sigma^2$ comparing with 
the corresponding coding RNAs. These results reveal distinct translation kinetics statistically between coding 
and noncoding RNAs, waiting biological significance of the findings to be discovered.

\section{Discussions}

We have applied stochastic simulation to study the translation kinetics at molecular level. In the model, RNA sequences, ribosome dynamics, tRNA pool and biochemical reactions in the elongation steps were included. Simulations showed that during translations the $ETC$ satisfied log-normal distribution (Fig. \ref{f:elong1}), and is mainly determined by both codon saturation (Fig. \ref{fig:elR}) and the steps of aa-tRNA selection (Fig. \ref{fig:sen}).  In the tRNA selection, the relative numbers of near-cognate to cognate aa-tRNAs are crucial for the elongation of a ribosome, and hence the mean value of the logarithm of  $ETC$ is dependent on the tRNA usages defined by $F_{\mathrm{cog}}$ and $F_{\mathrm{near}}$ in this paper (Fig. \ref{f:elong2}).  In the log-normal distribution Eq. \ref{eq:ga}, the mean $\mu$ and variance $\sigma^2$ are important for the density function of the elongation time. We showed that these two parameters are slightly different for coding and noncoding RNAs for both yeast and human samples (Fig. \ref{f:cnchuman}). On average, noncoding RNAs have smaller variance than coding RNAs. A simple model of ribosome dynamics revealed that the translation efficiency is mainly determined by the number of available ribosomes, translation initiation and the elongation time, and the translation efficiency depends on the elongation time through a Michaelis-Menten function. The translation efficiency increases with the available ribosomes number when the number is small, however is insensitive with the ribosome number when the number is large enough to saturate all codons.  These results are further confirmed by our simulations. Moreover, the translation accuracy decreases exponentially with sequence lengths.  These results suggest an improvement for the effective protein production when we are modeling gene expressions in gene regulation networks. 

In modeling gene expressions the protein production is described by a delay differential equation of form Eq. \ref{eq:P} 
that depends on the translation efficiency $\alpha$ and the distribution $\rho(\tau)$ of elongation time. This study 
showed that the effective production of correct proteins can be expressed as
\begin{equation}
\alpha = \dfrac{a e^{-c n}}{1 + b n e^{\mu + \sigma^2/2}},
\end{equation}  
here $n$ is the protein chain length (number of amino acids), $\mu$ and $\sigma^2$ are mean and variance of the 
logarithm of the $ETC$, respectively, parameters $a$ and $b$ are dependent on available ribosomes number,  
translation initiation and termination, and $c$ relates with the composition of the tRNA pool.  The distribution of the 
elongation time is formulated as 
\begin{equation}
\rho(\tau) = \dfrac{1}{\tau \sigma \sqrt{2 \pi}} e^{-\frac{(\ln \tau - \ln n - \mu)^2}{2\sigma^2}}.
\end{equation} 
Hence, the protein production equation Eq. \ref{eq:P} can be rewritten as
\begin{equation}
\label{eq:P2}
\dfrac{d P}{d t} = \dfrac{a e^{-c n}}{1 + b n e^{\mu + \sigma^2/2}} \int_0^{+\infty} M(t-\tau) \dfrac{1}{\tau \sigma \sqrt{2 \pi}} e^{-\frac{(\ln \tau - \ln n - \mu)^2}{2\sigma^2}} d \tau,
\end{equation}
where $M(t)$ is the number of mRNAs at time $t$. In this equation, the protein chain length $n$ is explicitly included. 
Moreover, the sequence information are implicitly included in the parameter $\tau$ and $\sigma^2$ which are mainly determined 
by the process of aa-tRNA selection in each step of ribosome movement. Yet other parameters are somehow universal under 
given cellular conditions. A direct conclusion from Eq. \ref{eq:P2} is the extreme low effective production rates of 
long proteins, which is because of the long elongation time and low translation accuracy for these long chain molecules.  
This is in consistent with biological observations that many transcription factors are small proteins with high 
production rates (many of them have high degradation rates as well) \cite{Babu:2004}, and many structural proteins (\textit{e.g.} fibers) 
and transport proteins (\textit{e.g.} membrane proteins) are large proteins with low production 
rates (these proteins are mostly very stable) \cite{Heijne:2006}. Hence, this study provides insightful details for the 
known observations, and is valuable in further works of whole cell modeling.

\begin{appendix}

\section*{Data Resources}
\label{}
RNA sequences are downloaded from available databases:
\begin{itemize}
\item Yeast coding RNAs from SGD (\small{http://downloads.yeastgenome.org/sequence/S288C\_reference/
orf\_dna/orf\_coding.fasta.gz}).
\item Yeast noncoding RNAs from SGD
(\small{http://downloads.yeastgenome.org/sequence/S288C\_reference/
rna/rna\_coding.fasta.gz}).
\item Human coding RNAs from Ensembl Genome Browser 
(\small{http://useast.ensembl.org/biomart/martview/8a921ac1ac4642b
07708af32f2339655}). 
\item Human noncoding RNAs from Genecode19 
(\small{ftp://ftp.sanger.ac.uk/pub/gencode/Gencode\_human/release\_19
/gencode.v19.lncRNA\_transcripts.fa.gz}).
\end{itemize}

\section*{Acknowledgement}
This work was supported by the National Natural Science Foundation of China (91430101 and 11272169).  We thank Prof. Zhi Lu and his lab members for valuable discussions.

\end{appendix}



  \bibliographystyle{authordate1}


\end{document}